\documentclass[twocolumn,tighten]{aastex631}
\usepackage{CJKutf8}
\usepackage{amsmath}
\usepackage{enumitem}
\usepackage{booktabs,pbox}
\usepackage{bm}

\graphicspath{{./figures/}{../figures/}}

\DeclareRobustCommand{\ion}[2]{%
\relax\ifmmode
\ifx\testbx\f@series
{\mathbf{#1\,\mathsc{#2}}}\else
{\mathrm{#1\,\mathsc{#2}}}\fi
\else\textup{#1\,{\mdseries\textsc{#2}}}%
\fi}

\newcommand{\hal}  {H$\alpha$}
\newcommand{\hb}   {H$\beta$}
\newcommand{\HI}   {\ion{H}{i}} 
\newcommand{\HII}  {\ion{H}{ii}} 
\newcommand{\OIII} {[\ion{O}{iii}]} 

\newcommand{\jwst} {JWST}
\newcommand{\hst}  {HST}
\newcommand{\m}    {$\mu$m}

\shorttitle{Clusters \& PAHs}
\shortauthors{Dale/PHANGS}

\DeclareUnicodeCharacter{2212}{-}
\begin{document}


\title{PAH Feature Ratios Around Stellar Clusters and Associations in 19 Nearby Galaxies}



\correspondingauthor{Daniel~A.~Dale}
\email{ddale@uwyo.edu}

\author[0000-0002-5782-9093]{Daniel~A.~Dale}
\affiliation{Department of Physics and Astronomy, University of Wyoming, Laramie, WY 82071, USA}
\author[0000-0002-5782-9093]{Gabrielle~B.~Graham}
\author[0000-0003-0410-4504]{Ashley~T.~Barnes}
\author[0000-0003-4974-3481]{Dalya~Baron}
\author[0000-0003-0166-9745]{Frank Bigiel}
\author[0000-0003-0946-6176]{M\'ed\'eric~Boquien}
\author[0000-0003-0085-4623]{Rupali~Chandar}
\author[0000-0002-5235-5589]{J\'er\'emy~Chastenet}
\author[0000-0001-8241-7704]{Ryan~Chown}
\author[0000-0002-4755-118X]{Oleg~V.~Egorov}
\author[0000-0001-6708-1317]{Simon C.~O.\ Glover}
\author{Lindsey~Hands}
\author[0000-0001-7448-1749]{Kiana~F.~Henny}
\author[0000-0002-4663-6827]{Remy Indebetouw}
\author[0000-0002-0560-3172]{Ralf S.\ Klessen}
\author[0000-0003-3917-6460]{Kirsten~L.~Larson}
\author[0000-0003-0946-6176]{Janice~C.~Lee}
\author[0000-0002-2545-1700]{Adam~K.~Leroy}
\author[0000-0001-6038-9511]{Daniel Maschmann}
\author[0000-0003-2721-487X]{Debosmita Pathak}
\author[0000-0002-0579-6613]{M.~Jimena~Rodríguez}
\author[0000-0002-5204-2259]{Erik~Rosolowsky}
\author[0000-0002-4378-8534]{Karin~Sandstrom}
\author[0000-0002-3933-7677]{Eva~Schinnerer}
\author[0000-0002-9183-8102]{Jessica~Sutter}
\author[0000-0002-8528-7340]{David~A.~Thilker}
\author[0009-0005-8923-558X]{Tony~D.~Weinbeck}
\author[0000-0002-3784-7032]{Bradley~C.~Whitmore}
\author[0000-0002-0012-2142]{Thomas~G.~Williams}
\author[0000-0001-8289-3428]{Aida Wofford}


\begin{abstract}
We present a comparison of observed polycyclic aromatic hydrocarbon (PAH) feature ratios in 19 nearby galaxies with a grid of theoretical expectations for near- and mid-infrared dust emission.  The PAH feature ratios are drawn from Cycle~1 JWST observations and are measured for 7\,224 stellar clusters and 29\,176 stellar associations for which we have robust ages and mass estimates from HST five-band photometry.  Though there are galaxy-to-galaxy variations, the observed PAH feature ratios largely agree with the theoretical models, particularly those that are skewed toward more ionized and larger PAH size distributions.  For each galaxy we also extract PAH feature ratios for 200~pc-wide circular regions in the diffuse interstellar medium, which serve as a non-cluster/association control sample.  Compared to what we find for stellar clusters and associations, the 3.3\m/7.7\m\ and 3.3\m/11.3\m\ ratios from the diffuse interstellar medium are $\sim 0.10-0.15$~dex smaller.  When the observed PAH feature ratios are compared to the radiation field hardness as probed by the \OIII/\hb\ ratio, we find anti-correlations for nearly all galaxies in the sample.  These results together suggest that the PAH feature ratios are driven by the shape intensity of the radiation field, and that the smallest PAHs---observed via JWST F335M imaging---are increasingly `processed' or destroyed in regions with the most intense and hard radiation fields.


\end{abstract}




\section{Introduction} \label{sec:intro}
Polycyclic aromatic hydrocarbons (PAHs) are ubiquitous in the interstellar medium of galaxies \citep{draine2011,tielens2010}, and their conspicuous emission features at near- and mid-infrared wavelengths can be leveraged as a useful probe of the incident radiation field arising from both nearby luminous sources and the general interstellar radiation field.  Theoretical work on dust infrared emission indicates that varying the shape (hardness) of the radiation field can lead to changes in the ratios of infrared PAH feature emission, for both neutral and ionized PAHs (e.g., \citealt{lidraine2001}).  The intensity of the radiation field is another parameter that can impact PAH feature ratios; according to the models of \cite{draine2021} (see their Figure~19), changes in the radiation field intensity leave PAH feature ratios relatively unaltered until the intensities exceed $\sim 10^3$ times that of the local interstellar radiation field, a regime for which the 3.3\m/7.7\m\ and 3.3\m/11.3\m\ PAH ratios increase.  The ionization of the smallest dust grains can also impact their emission; the prominent 7.7\m\ PAH emission is expected to be particularly strong for moderate levels of ionization, whereas the 3.3\m\ and 11.3\m\ PAH emission features are weaker for higher ionization levels.  Similarly, the relative strengths of PAH emission features depend on their sizes---larger PAH size distributions lead to relatively weak 3.3\m\ emission compared to the emission from PAHs at longer infrared wavelengths \citep{maragkoudakis2020}.  This trend may be understood in terms of the relative heat capacities for PAHs: smaller PAHs will experience higher peak temperatures after absorbing a single UV photon, resulting in higher 3.3\m/7.7\m\ and 3.3\m/11.3\m\ PAH ratios for a given radiation field.  Laboratory work also has shown that the relative strengths of the PAH emission features can depend on grain size \citep{allamondola1999,shannon2019,maragkoudakis2023}.  In short, the collective effort from both theory and laboratory experiments suggest that studying PAH feature ratios can lead to sensitive constraints on their physical properties along with the properties of the radiation fields that heat these small dust grains.  

Detailed observational work on the relative shapes and strengths of the infrared PAH features has been ongoing since the {\it Infrared Space Observatory} and {\it Spitzer Space Telescope} missions \citep{cesarsky1996,madden2006,spoon2007,smith2007,dale2009,diamondstanic2010,sandstrom2010,sandstrom2012,wu2010} and continues today with Akari and some ground-based efforts \citep{lai2020,ramosalmeida2023,kondo2024}.  However, the exquisite angular resolution and sensitivity of \jwst\ has broken new ground on PAH studies---the 3.3\m\ PAH feature is now routinely detected and can be mapped on pc scales in extragalactic systems and on much smaller scales within the Galaxy \citep{lai2023,chown2023,bolatto2024,milisavljevic2024,pedrini2024,peeters2024}.  The Physics at High Angular resolution in Nearby GalaxieS (PHANGS) project \citep{leroy2021,emsellem2022,lee2022,lee2023} is obtaining comprehensive \jwst\ NIRCam and MIRI imaging in several medium and wide bands for 74 nearby star-forming disk galaxies (Cycle~1 GO~2107; Cycle~2 GO~3707).  The Cycle~1 PHANGS imaging dataset composed of the F335M, F770W, and F1130W bands targets the PAH emission features at 3.3, 7.7, and 11.3\m.  Initial PHANGS efforts that focused on the near- and mid-infrared PAH emission include: a prescription for producing continuum-subtracted 3.3\m\ PAH maps \citep{sandstrom2023}, employing PAHs to probe the hardness and/or intensity of the radiation field \citep{egorov2023,baron2024}, mapping the structure and properties of the interstellar medium \citep{chastenet2023a,chastenet2023b,leroy2023,meidt2023,thilker2023,pathak2024}, using PAH emission to trace embedded star clusters and star-forming regions \citep{kim2023,rodriguez2023,schinnerer2023}, and PAHs as tracers of galactic ``bubbles'' \citep{barnes2023,watkins2023}.  

\cite{dale2023} compared PAH feature ratios for 1\,063 stellar clusters and 2\,654 stellar associations within the first three PHANGS galaxies for which the full suite of NIRCam and MIRI imaging was available: NGC~0628, NGC~1365, and NGC~7496 (\citealt{dale2023}; hereafter Paper~I).  When compared against recent theoretical dust emission spectra, this preliminary work showed that the PAH populations near stellar clusters and associations in these three galaxies were skewed to larger sizes and higher ionization levels.  The results presented here in ``Paper~II'' build upon these prior efforts by expanding the dataset using the full sample of 19 galaxies that were imaged in the PHANGS Cycle~1 Treasury program (PID 2107; PI J. Lee).  In addition, we expand the analysis to include PAH feature ratios extracted from diffuse regions spread across each galaxy disk.

Section~\ref{sec:sample_data} provides an overview of the sample and observational data. Section~\ref{sec:analysis} reviews the approaches taken to analyze the data and how theoretical models of dust emission are folded into the analysis.  Section~\ref{sec:results} presents the results, Section~\ref{sec:discussion} interprets the results within the context of the current literature, and Section~\ref{sec:conclusions} summarizes the main findings.


\section{Sample and Data} \label{sec:sample_data}
\begin{figure*}[!t]
 \includegraphics[height=18cm,trim={0 0 0 0},clip]{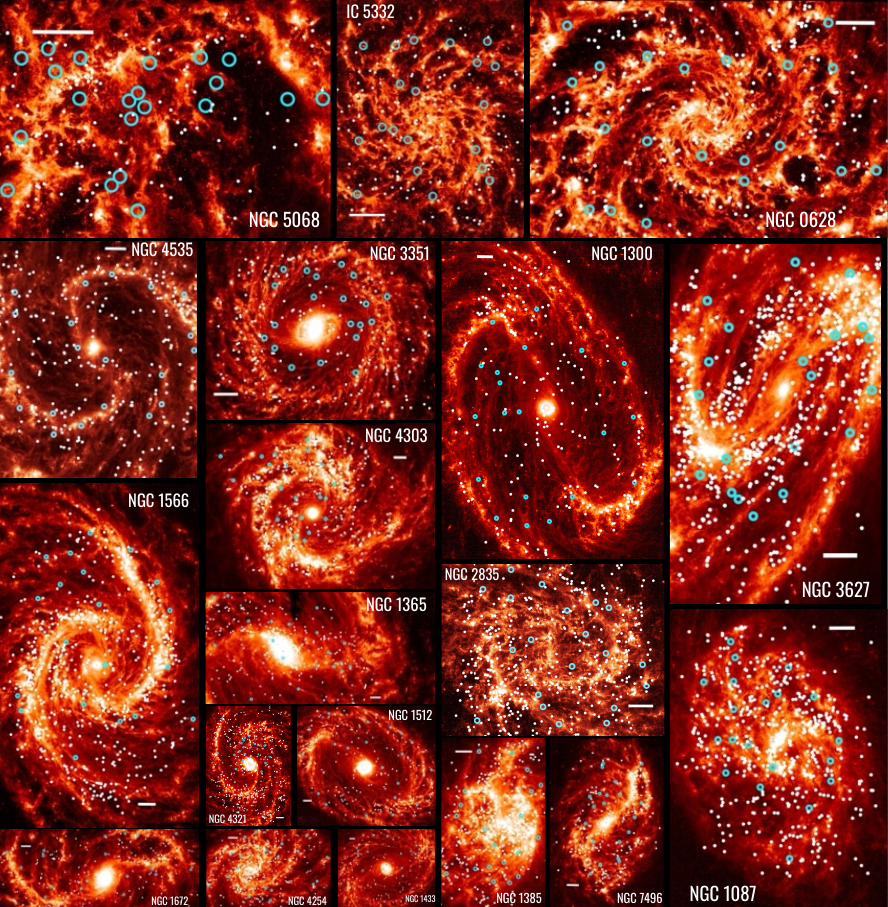}
 \caption{All nineteen Cycle~1 PHANGS-JWST galaxies as seen by the JWST/MIRI F770W filter.  Small white dots indicate the locations of stellar clusters for which we study PAH feature ratios.  Large blue circles indicate 200~pc diameter diffuse regions that do not overlap with stellar clusters, stellar associations, or \HII\ regions and serve as a control sample in the analysis.  Horizontal scale bars indicate 1~kpc.}
 \label{fig:all}
\end{figure*}

The galaxies and observations analyzed here are drawn from the PHANGS JWST Cycle~1 Treasury project.  The 19 galaxies studied here are listed in Table~\ref{tab:sample}.  All 19 targets are $\sim$solar metallicity, star-forming main sequence galaxies but six are known to host Seyfert nuclei (see Table~\ref{tab:sample}).  Thirteen systems show prominent spiral arms and 15 exhibit well-defined stellar bars \citep{querejeta2021}.  The four NIRCam mosaics obtained for each galaxy employ the F200W, F300M, F335M, and F360M filters and the four MIRI mosaics use the F770W, F1000W, F1130W, and F2100W filters.  In this study, we utilize all of the imaging data except for those taken with the F2100W filter.  Figure~\ref{fig:all} provides our F770W mosaics along with overlays for the stellar clusters and diffuse regions described below.

\begin{deluxetable}{lccrrr}
\tablenum{1}
\tablecaption{Galaxy Sample}
\tablewidth{0pc}
\label{tab:sample}
\tablehead{
\colhead{Galaxy} & \colhead{R.A.}  & \colhead{Dec.}  & \colhead{~$D$~} & \colhead{~$M_*$~}   & \colhead{log$_{10}$SFR}\\
\colhead{}       & \colhead{J2000} & \colhead{J2000} & \colhead{Mpc}   & \colhead{$M_\odot$} & \colhead{$M_\odot$/yr}}
\startdata
NGC~0628          & 01:36:41.7 & $+$15:47:01 & ~9.84 & 10.3&0.24\\ 
NGC~1087          & 02:46:25.2 & $-$00:29:55 & 15.85 &  9.9&0.11\\ 
NGC~1300          & 03:19:41.0 & $-$19:24:40 & 18.99 & 10.6&0.07\\ 
NGC~1365$\dagger$ & 03:33:36.4 & $-$36:08:25 & 19.57 & 11.0&1.24\\ 
NGC~1385          & 03:37:28.6 & $-$24:30:04 & 17.22 & 10.0&0.32\\ 
NGC~1433          & 03:42:01.5 & $-$47:13:19 & 18.63 & 10.9&0.05\\ 
NGC~1512          & 04:03:54.1 & $-$43:20:55 & 18.83 & 10.7&0.11\\ 
NGC~1566$\dagger$ & 04:20:00.4 & $-$54:56:17 & 17.69 & 10.8&0.66\\ 
NGC~1672$\dagger$ & 04:45:42.5 & $-$59:14:50 & 19.40 & 10.7&0.88\\ 
NGC~2835          & 09:17:52.9 & $-$22:21:17 & 12.22 & 10.0&0.10\\ 
NGC~3351          & 10:43:57.8 & $+$11:42:13 & ~9.96 & 10.4&0.12\\ 
NGC~3627$\dagger$ & 11:20:15.0 & $+$12:59:29 & 11.32 & 10.8&0.59\\ 
NGC~4254          & 12:18:49.6 & $+$14:25:00 & 13.10 & 10.4&0.49\\ 
NGC~4303$\dagger$ & 12:21:54.9 & $+$04:28:25 & 16.99 & 10.5&0.73\\ 
NGC~4321          & 12:22:54.9 & $+$15:49:20 & 15.21 & 10.7&0.55\\ 
NGC~4535          & 12:34:20.3 & $+$08:11:53 & 15.77 & 10.5&0.34\\ 
NGC~5068          & 13:18:54.7 & $-$21:02:19 & ~5.20 &  9.4&$-$0.56\\ 
NGC~7496$\dagger$ & 23:09:47.3 & $-$43:25:40 & 18.72 & 10.0&0.35\\ 
IC~5332           & 23:34:27.5 & $-$36:06:04 & ~9.01 &  9.7&$-$0.39\\ %
\enddata
\tablenotetext{}{Values are from \cite{leroy2019} and references therein.  $\dagger$ indicates a Seyfert classification from \cite{veroncetty2010}, though for NGC~3627 recent evidence suggests that its central emission is more consistent with ionization by hot low-mass evolved stars \citep{belfiore2022}.}
\end{deluxetable}

An initial description of the observations and the data processing for this JWST program were presented in \cite{lee2023}, and \cite{williams2024} provides an update to the processing (\texttt{pjpipe}) based on several recent implementations that go beyond the standard pipeline procedures.  The enhancements include optimized relative and absolute astrometry that leverages red point sources (candidates for asymptotic giant branch and other red evolved stars) identified in HST imaging; improved anchoring of the flux levels using archival Spitzer/IRAC and WISE imaging; background matching between tiles in a mosaic; background subtraction for MIRI imaging; and better removal of NIRCam instrumental artifacts such as 1/$f$ noise.  The images analyzed here are from Version~1.1 of the \texttt{pjpipe} data processing and are available via the MAST archive.\footnote{https://archive.stsci.edu/hlsp/phangs/phangs-jwst}

The photometry described in the next section focuses on regions centered on stellar clusters and associations from the PHANGS--\hst\ program \citep{lee2022,thilker2022,larson2023,maschmann24}.  We restrict our cluster-based analysis to the compact Class~1 and 2 types \citep{whitmore2021,deger2022} and the stellar association-based analysis relies on the 32~pc-scale $V$-band catalog, as described in \cite{larson2023}. The estimation of ages and masses for the clusters and associations is based on spectral energy distribution fits to $NUV$$-$$B$$-$$V$$-$$R$$-$$I$ \hst\ datasets \citep{turner2021,henny2025,thilker2025}.  For the 19 galaxies studied here there are a total of $7\,224$ compact clusters and $29\,176$ stellar associations in the publicly-available catalogs that overlap with the PHANGS--\jwst\ footprints.  To explore a wider variety of environments in these galaxies, we also identified by eye 20 additional regions per galaxy that are centered on diffuse regions of the interstellar medium that do not overlap with cataloged stellar clusters \citep{turner2021}, stellar associations \citep{larson2023}, or \HII\ regions \citep{belfiore2022,santoro2022,congiu2023,groves2023}.  These diffuse regions are meant to serve as a control sample for comparison against the analysis carried out for stellar clusters and associations.  Additional photometric studies of PAH emission in PHANGS galaxies extend our work here to environments such as those dominated by spiral arms, circumnuclear rings, CO emission, \HI\ emission, or simply all areas outside of known \HII\ regions (e.g., \citealt{chastenet2023a,chastenet2023b,egorov2023,baron2024,sutter2024}).

For Paper~I we removed clusters and associations from our analysis that spatially overlapped with saturated regions of the NIRCam/MIRI imaging associated with the active galactic nuclei for NGC~1365 and NGC~7496.  For the additional 16 PHANGS galaxies studied here in Paper~II, the only other galaxy exhibiting saturated pixels in the NIRCam/MIRI imaging for the wavelengths studied here is NGC~1566, another PHANGS galaxy known to host a Seyfert nucleus \citep{veroncetty2010}.  For this galaxy we removed the 12 stellar associations lying closest to the nucleus (none of the stellar clusters lies in the saturated zone); none of our diffuse regions or stellar clusters overlapped with saturated regions in NGC~1566.


\section{Analysis} \label{sec:analysis}
Our aim is to compare the relative strengths of the 3.3, 7.7, and 11.3\m\ PAH emission features with theoretical expectations from dust models.  As such, we restrict our analysis to the PAH emission features that are captured by the JWST F335M, F770W, and F1130W bandpasses.  However, in addition to using the imaging that is based on these three filters, we also use the JWST F300M and F360M imaging in conjunction with the recipe outlined in \cite{sandstrom2023} in order to subtract off the underlying stellar and hot dust continuum detected by the F335M filter.  We refer to this continuum-subtracted 3.3\m\ flux as ``F335M$_{\rm PAH}$''.  Likewise, we adopt the approach of \cite{sutter2024} and use a fraction of the F200W flux to remove the stellar continuum contribution at 7.7\m; we use their stellar fraction of F770W$_*$/F200W = 0.13$\pm$0.04.  The stellar continuum contribution to F1130W is minimal and no correction is made for the hot dust continuum to F770W or F1130W (see \citealt{baron2024b}).\footnote{We show in the Appendix that removing the residual hot dust continuum for the F770W and F1130W bands does not significantly impact our results or change our conclusions.}  Before attempting any image analysis, all images besides the F1130W images are first convolved with appropriate smoothing kernels\footnote{https://github.com/francbelf/jwst\_kernels} in order to generate smoothed images with resolution given by the F1130W PSF, which has the lowest resolution in our study  ($\approx$0\farcs38 full-width-half-maximum; see \citealt{williams2024} for details).

Photometric extractions at each wavelength for each stellar cluster were carried out using 0\farcs6 diameter circular apertures.\footnote{No aperture corrections are necessary since we only analyze ratios of fluxes derived from imaging at the same resolution.}  For our galaxy sample with distances of 5--20~Mpc, 0\farcs6 diameters correspond to 15--57~pc.  Photometry was likewise carried out at each wavelength for each stellar association using the polygons defined in \cite{larson2023} which derive from a watershed analysis that is based on $V$-band point-source detections and a 32~pc FWHM Gaussian smoothing.  The typical polygon for this sample has an effective diameter---defined by the area of the effective circle that matches that of the polygon---of $\sim$1\farcs1 which translates to $\sim$55~pc for our typical galaxy distance of 10~Mpc.  To help promote sufficient signal-to-noise ($>5$) values for the diffuse regions which by their nature are typically much fainter than star-forming regions at NIRCam and MIRI wavelengths, particularly for the NIRCam observations, we use a fixed 200~pc diameter which spans 2\farcs1--8\farcs0 for the range of distances of these 19 galaxies.  No local background subtractions were employed.  Tests show that if local backgrounds are subtracted, the results in Section~\ref{sec:results} are qualitatively the same but with increased scatter due to lower signal-to-noise values on average.

For a cluster, association, or diffuse region to be included in the analysis, we required a signal-to-noise of at least 5 (after continuum subtraction) for each of the NIRCam and MIRI bands used here.  The noise level for each image was determined by taking the standard deviation of the median value in at least 50 0\farcs6 diameter apertures scattered throughout the outer portions of the image that did not have obvious contamination from the galaxy or any foreground or background sources.
This signal-to-noise$>$5 requirement for all bands reduces the number of clusters available for analysis from 7\,224 to 6\,031, the number of associations from 29\,176 to 24\,634, and the number of diffuse regions from 380 to 287.

To be consistent with the analysis for Paper~I, we rely on the models of \cite{draine2021} in order to compare the observed PAH feature ratios to theoretical expectations.  To enable fair comparisons between the observations and the models, we \textbf{have implemented the aforementioned stellar+dust+PAH emission models} into the CIGALE \citep{boquien2019} \textbf{spectro-photometric modeling code, which we have used to compute} synthetic fluxes using the known JWST filter bandpasses.  The simulated spectra for which we extract synthetic spectra are described by: PAH ionizations spanning $\texttt{ion}=0,1,2$ (``low'', ``standard'', and ``high''; see \citealt{draine2021}, their Figure~9b), PAH size distributions covering $\texttt{size}=0,1,2$ corresponding to $a_{01}=3,4,5~$\AA\ (see \citealt{draine2021}, their Figure~9a), solar metallicity \cite{bruzual2003} stellar populations with a standard and fully sampled \cite{chabrier2003} initial mass function, interstellar radiation field intensities $U$ ranging over $\texttt{logU}=0-7$ where $U=1$ is approximately the radiation field in the Solar Neighborhood, and ages of the stellar populations that are heating the dust having values of $\texttt{age}=3,10,100,1000$~Myr.  More precisely, we assume an instantaneous burst of star formation and thus these stellar population ages are the times that have elapsed since the burst.  With this approach for an aging stellar population we effectively vary the hardness of the radiation field.  Nonetheless, we emphasize that the \cite{draine2021} models do not tie the ``processing'' of the dust grains to any environmental parameter; parameters such as grain ionization and grain size are possible for any chosen combination of interstellar radiation field intensity, stellar population age, and metallicity.  Finally, we note that the \cite{draine2021} models, unlike previous iterations of the Draine dust models, assume a fixed PAH dust mass fraction and no longer invoke a $\gamma$ parameter that distinguishes between dust heated by a distribution of radiation field intensities and diffuse dust heated by a minimum radiation field intensity.  See Figure~2 of Paper~I for a depiction of six example model spectra.  We have applied to our synthetic data the prescription outlined in \cite{sandstrom2023} for removing the underlying hot dust plus stellar continua to the 3.3\m\ PAH emission and the \cite{sutter2024} prescription for removing the underlying stellar continuum to the 7.7\m\ PAH feature emission consistent with what is done for the observations.


\section{Results} \label{sec:results}
\begin{figure}
 \includegraphics[height=16.5cm]{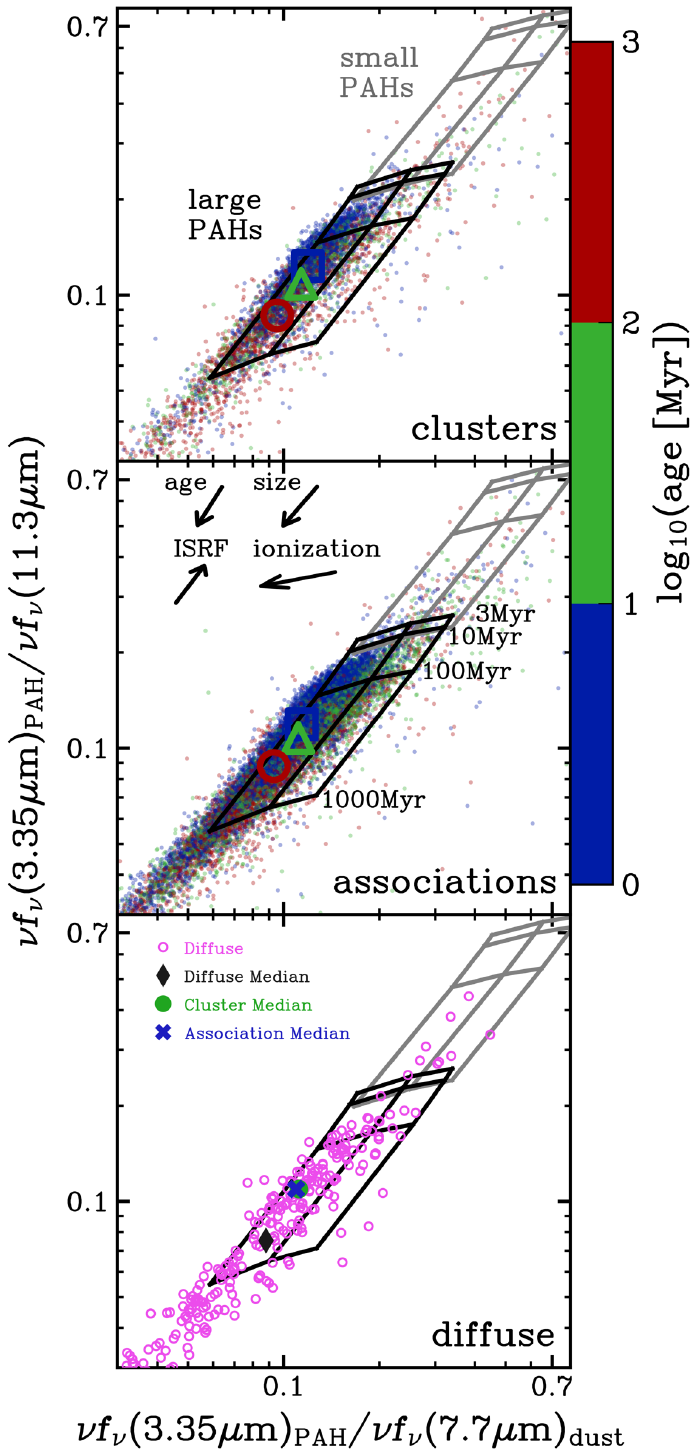}
 \caption{Top: Continuum-subtracted flux ratios for stellar clusters in all 19 PHANGS-JWST Cycle~1 galaxies.  The small data points are colored according to their ages as quantified in the color bar.  The large open symbols indicate the median values for each age bin.  Overlaid tracks are included for select subsets of the synthetic ratios extracted from the dust emission models of \cite{draine2021}, and the general trends for these tracks are indicated with the inset arrows/descriptions.  The grey and black tracks are for small ($a_{01}=3~$\AA) and large ($a_{01}=5~$\AA) modeled PAH distributions, respectively, with both assuming \texttt{logU}=0, \texttt{ion}=0,1,2, and \texttt{age}=3,10,100,1000~Myr.  Middle: Using stellar associations instead of compact clusters.  Bottom:  Individual diffuse regions along with the median values for clusters, associations, and diffuse regions.}
 \label{fig:ratios.one.panel}
\end{figure}

Figure~\ref{fig:ratios.one.panel} provides the F335M$_{\rm PAH}$/F1130W and F335M$_{\rm PAH}$/F770W$_{\rm dust}$ feature ratios for the stellar clusters and associations in all 19 JWST-PHANGS Cycle~1 galaxies.  Included in Figure~\ref{fig:ratios.one.panel} is the grid of synthetic points from the models of \cite{draine2021} described in \S~\ref{sec:analysis}.  The black and grey grids in Figure~\ref{fig:ratios.one.panel} demonstrate how the synthetic points depend on stellar cluster age (i.e., the shape of the stellar cluster spectrum), PAH size distribution, and PAH ionization for a fixed interstellar radiation field (ISRF) intensity $U=1$.  The large arrows and corresponding descriptors indicate how the synthetic ratios change for each modeled parameter.  The 7.7\m\ PAH emission is particularly sensitive to the ionization level (e.g., Figure~2 of \citealt{lidraine2001}), and thus the ionization arrow in Figure~\ref{fig:ratios.one.panel} is mostly horizontal.  The trends for the other parameters largely follow the orientation of the synthetic grids.  For example, the 3.3\m/7.7\m\ and 3.3\m/11.3\m\ ratios are predicted to decrease for increasing ages of the stellar population that is responsible for heating the interstellar dust and PAHs, because the softer radiation fields involved lead to lower peak dust temperatures (e.g., Section~9.7 of \citealt{draine2021}).  The overall distributions of the observed ratios in Figure~\ref{fig:ratios.one.panel} track well with the grid of synthetic models, but there is an appreciable portion of the dataset that falls below the displayed grids.  \cite{draine2021} acknowledge that their synthetic 3.3\m/7.7\m\ and 3.3\m/11.3\m\ ratios are noticeably larger than observations of normal star-forming galaxies suggest, at least for typical PAH distributions, ionization levels, and interstellar radiation fields.  For the observed PHANGS--JWST ratios that fall within the synthetic grids, the majority are consistent with the PAHs having elevated ionization levels and large size distributions.  In other words, the data are mostly populating the black grid for larger PAHs (and not the grey grid for smaller PAHs) between the ``standard'' and ``high'' ionization model tracks forming the left half of the black grid.  For both clusters and associations, the youngest clusters (1--10~Myr) are on average marginally closer to the highest ionization levels.

\begin{figure*}
 \includegraphics[height=15.5cm,clip]{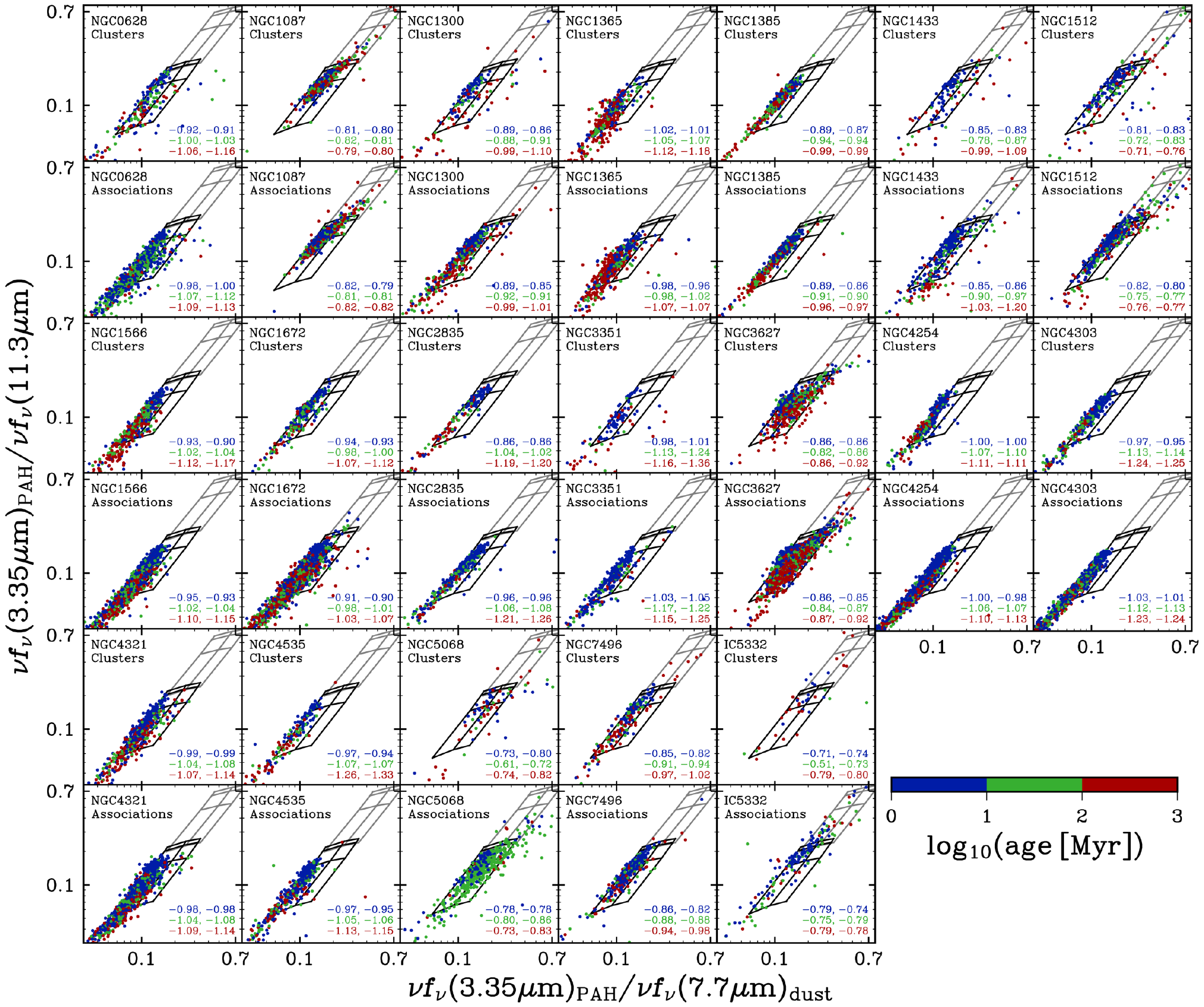}
 \caption{The data from Figure~\ref{fig:ratios.one.panel} expanded into separate panels for each galaxy.  Logarithmic medians are presented within each panel.  While there are galaxy-to-galaxy differences, the overall sample shows no significant differences between the medians and dispersions for clusters and associations.}
 \label{fig:ratios.multi.panel}
\end{figure*}

Figure~\ref{fig:ratios.multi.panel} expands the data from Figure~\ref{fig:ratios.one.panel} into separate portrayals for each galaxy.  Two panels are provided for each galaxy, with the feature ratios for stellar clusters (associations) appearing in the top (bottom) panel.  Table~\ref{tab:medians} lists the median values and the statistical spreads for each galaxy's distributions in Figure~\ref{fig:ratios.multi.panel}.  There are no significant differences between the ensemble medians and dispersions for the clusters compared to the associations (Figure~\ref{fig:ratios.one.panel}), but there are some galaxy-to-galaxy differences evident in Figure~\ref{fig:ratios.multi.panel} and Table~\ref{tab:medians}.  For example, most clusters and associations for NGC~1512 and IC~5332 in Figure~\ref{fig:ratios.multi.panel} are skewed higher up the black (large PAH) grid, suggestive of dust heating by younger/harder radiation fields.  Conversely, nearly all of the data points for NGC~1365, NGC~4254, and NGC~4303 are shifted down the grid and lie between black model grid lines corresponding to effective stellar population ages of 100~Myr--1~Gyr, indicative of older and/or softer-than-average radiation fields.  

To further explore the dependence of PAH strength on the characteristics of the radiation field impinging on the dust grain population, the bottom panel of Figure~\ref{fig:ratios.one.panel} shows the PAH feature ratios for the selected diffuse regions described in Section~\ref{sec:sample_data}.  In terms of ionization, we find no discernible difference for the PAH ratio distribution for diffuse regions with respect to the clusters and associations.  However, the diffuse data points appear to be shifted ``down'' parallel to the grid.  This downward shift can be interpreted as representing radiation fields that are older/softer/less intense and/or larger PAH size distributions.  For reference, \cite{baron2024} found that the 3.3\m/11.3\m\ ratio is about 0.1~dex lower for regions characterized by diffuse ionized gas compared to star-forming regions, similar in amplitude to the shift we see for diffuse regions compared to regions centered on stellar clusters and associations.

\begin{figure*}
 \includegraphics[height=7.5cm,clip]{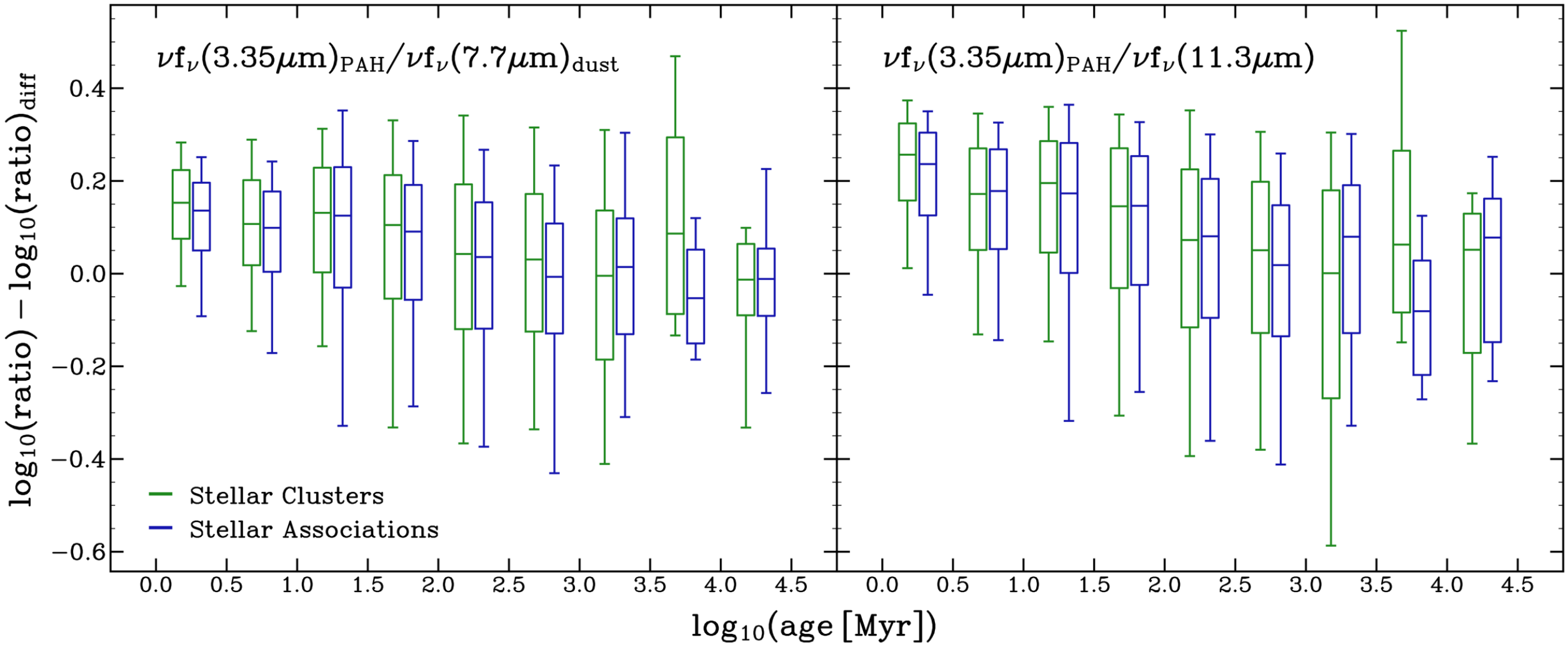}
 \caption{A box-whisker plot of the data from Figure~\ref{fig:ratios.one.panel} as a function of stellar cluster age.  The $y$-axis values for each age bin are the median ratios normalized by the median diffuse value.  The horizontal bars within each box indicate the median values, the boxes themselves span the 25$^{\rm th}$--75$^{\rm th}$ percentiles, and the whiskers span the 10$^{\rm th}$--90$^{\rm th}$ percentiles.  Linear fits to the data have Spearman's rank correlation coefficients ranging between $-$0.833 and $-$0.950 and two-sided significance levels of their deviations from zero between 0.001 and 0.005.}
 \label{fig:ratios.vs.ages}
\end{figure*}

In Figure~\ref{fig:ratios.vs.ages} we plot the median PAH feature ratios, normalized by the diffuse control sample's median ratio, as a function of binned stellar cluster age.  We see in Figure~\ref{fig:ratios.vs.ages} a convincing anti-correlation between the PAH feature ratios and cluster age in agreement with predictions from theoretical dust models \citep[e.g., Figure~16 of][]{draine2021}.

\begin{figure}
 \includegraphics[height=11.5cm,clip]{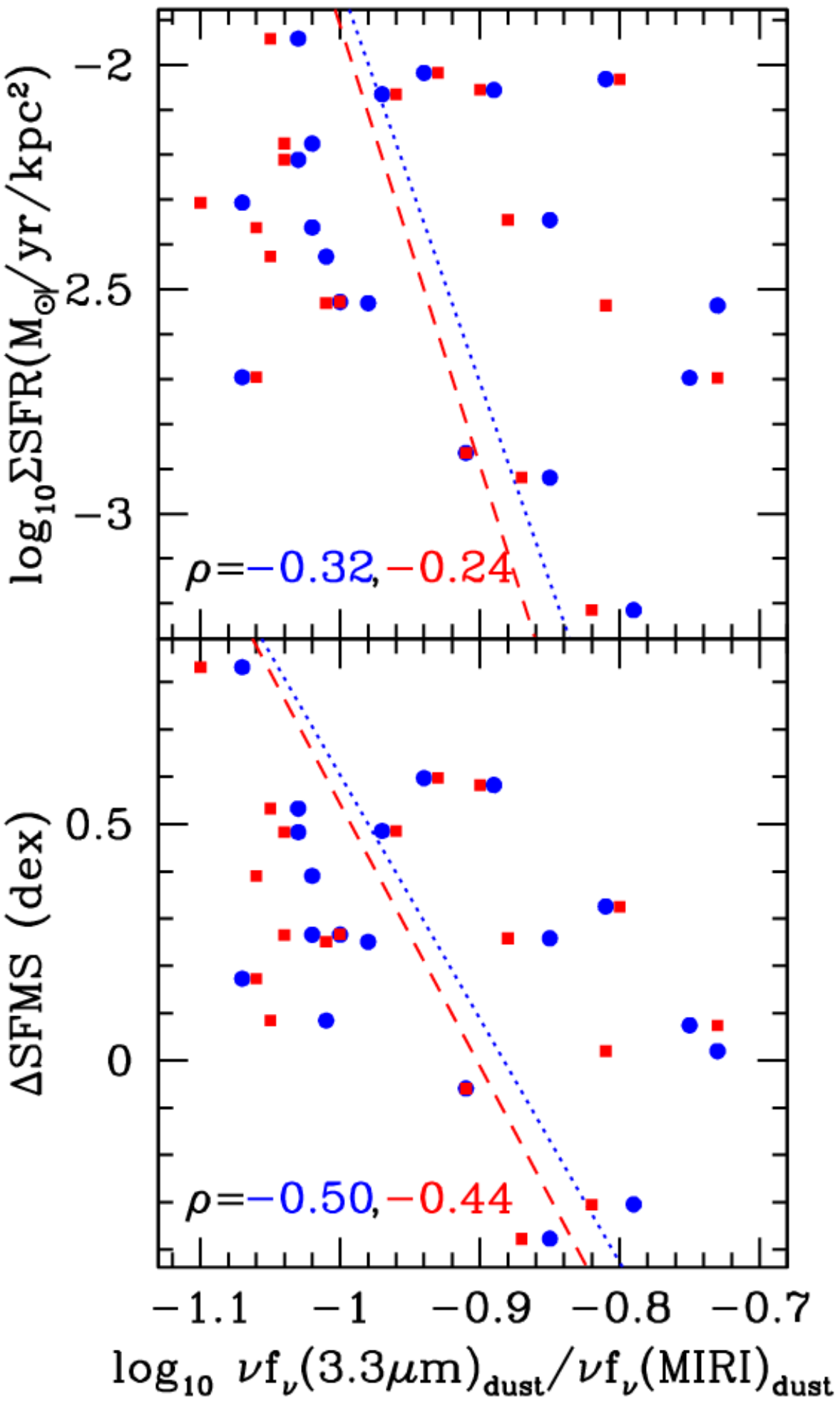}
 \caption{Average star formation rate surface densities (top) and spatially-integrated offsets from the star-forming main sequence (bottom) as a function of median PAH feature ratios for each galaxy's clusters.  Blue circles (red squares) use the F770W (F1130W) data.  The best-fit trend lines are included along with the Spearman's rank correlation coefficients.  The coefficients and the two-sided significance levels of their deviations from zero are ($-$0.33,0.17), ($-$0.25,0.29), ($-$0.50,0.03), and ($-$0.45,0.05).}
 \label{fig:offsets}
\end{figure}

Since the stellar clusters and associations follow the expectations for ionized-and-large PAHs in Figure~\ref{fig:ratios.multi.panel}, then the remaining two physical parameters we have explored with the synthetic grids---the shape and intensity of the radiation field---would imply that F335M$_{\rm PAH}$/F1130W and F335M$_{\rm PAH}$/F770W$_{\rm dust}$ feature ratios would both increase for larger values of the disk-averaged star formation rate surface densities ($\Sigma_{\rm SFR}$); as explained in \S~\ref{sec:intro}, both of these ratios are predicted by the \cite{draine2021} models to increase for higher values of the ISRF and for harder radiation fields.  Interestingly, the data for this sample seem to suggest the opposite result.  For example, NGC~4254 has the highest $\Sigma_{\rm SFR}$ of these 19 galaxies (0.011~$M_\odot$~yr$^{-1}$~kpc$^{-2}$) and NGC~1512 the lowest (0.00061~$M_\odot$~yr$^{-1}$~kpc$^{-2}$), but the PAH feature ratios in Figure~\ref{fig:ratios.multi.panel} for NGC~4254 are relatively low and the PAH feature ratios for NGC~1512 are relatively high.  Figure~\ref{fig:offsets} plots for the entire 19 galaxy Cycle~1 sample the galaxy-averaged $\Sigma_{\rm SFR}$ values in the top panel (see also \citealt{ujjwal2024}) and galaxy-integrated offsets from the star-forming main sequence \citep{leroy2021} in the bottom panel as a function of galaxy-averaged F335M$_{\rm PAH}$/F1130W and F335M$_{\rm PAH}$/F770W$_{\rm dust}$ values.  The trend described anecdotally with NGC~4254 and NGC~1512 extends to the full 19 galaxy sample.

One possible explanation for this contradictory result hinges on the lack of ``processing'' of PAHs in the \cite{draine2021} models.  The ionization levels of the PAHs are not tied to the characteristics of the radiation field, and there is no PAH destruction built into the models.  However, it has been well established that PAH emission is weak in regions of vigorous star formation associated with younger stellar ages and intense radiation fields \citep{cesarsky1996,madden2006,dale2009,sandstrom2012,galliano2021,egorov2023}.  The 3.3\m\ feature traces the smallest PAHs, and thus may be the most likely PAH feature to appear weak for intense and/or hard radiation fields due to photo-evaporation, sputtering, photo-desorption, etc.\ \citep{galliano2022}.

\begin{figure}
 \includegraphics[height=8cm,clip]{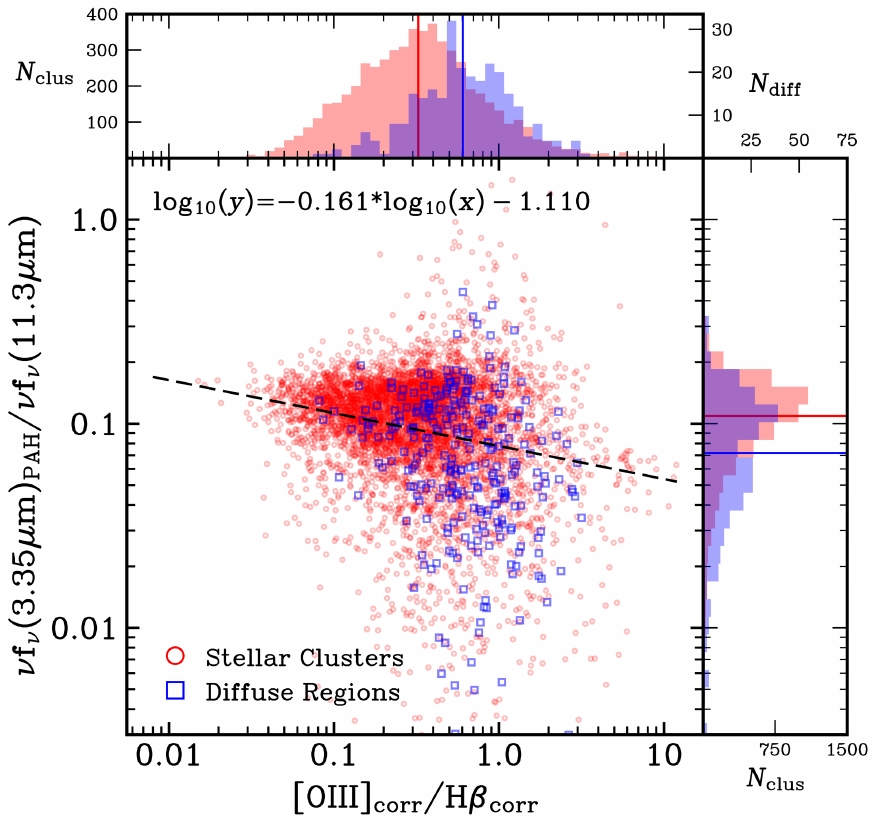}
 \includegraphics[height=8cm,clip]{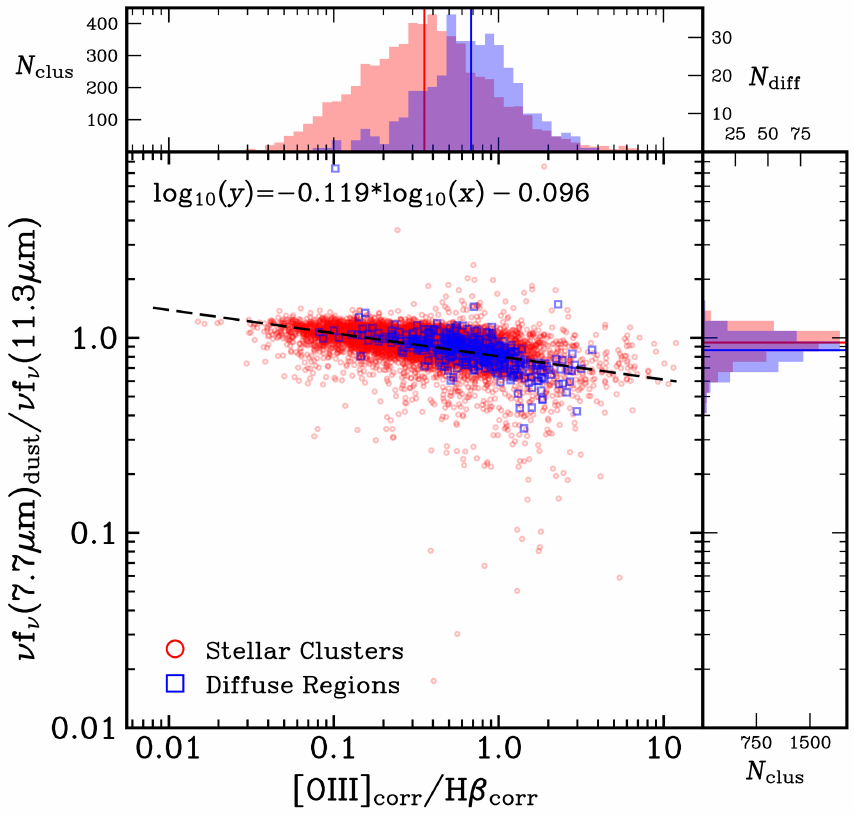}
 \caption{The F335M$_{\rm PAH}$/F1130W (top) and F770W$_{\rm dust}$/F1130W (bottom) ratios as a function of radiation field hardness as measured by the \OIII/\hb\ ratio corrected for extinction.  The ratios for stellar clusters are shown as red circles and red histograms, and for the diffuse regions as blue squares and blue histograms.  The dashed black line shows the linear fit to the combined clusters$+$diffuse regions.  The lines for the histograms indicate the median values.}
 \label{fig:hardness}
\end{figure}

To explore whether there are any dependencies on the hardness of the radiation field, we compare the F335M$_{\rm PAH}$/F1130W and F770W$_{\rm dust}$/F1130W ratios as a function of \OIII/\hb, a proxy for radiation field hardness \citep[e.g.,][]{egorov2023}.  The \OIII5007/\hb\ ratios are taken from the PHANGS--MUSE IFU spectral mapping program described by \cite{emsellem2022}. For this application we utilize the `copt' (convolved optimized) MUSE mosaics, which derive from datacubes with point spread functions (PSFs) that have been homogenized across the entire wavelength range and spatial extent of the mosaic on a galaxy-by-galaxy basis.  The `copt' PSFs for these 19 galaxies range from 0\farcs56 to 1\farcs25.  For comparison, the photometric apertures described in Section~\ref{sec:analysis} are 0\farcs6, 1\farcs1, and 2\farcs1--8\farcs0, respectively, for stellar clusters, stellar associations, and diffuse regions.  The \OIII\ and \hb\ fluxes have been corrected for extinction intrinsic to the external galaxy using the observed Balmer decrement and the \cite{calzetti2000} reddening curve assuming that the intrinsic \hal/\hb\ ratio is 2.86.  The data for stellar clusters and diffuse regions from all 19 galaxies are shown in Figure~\ref{fig:hardness}, though we note that the four lowest-mass galaxies in our sample individually show no trend in F335M$_{\rm PAH}$/F1130W versus \OIII5007/\hb\ (NGC~1087, NGC~2835, NGC~5068, and IC~5332) and the lowest mass galaxy shows no trend in F770W$_{\rm dust}$/F1130W versus \OIII5007/\hb\ (NGC~5068).  Though there is substantial scatter in the distribution of the data plotted in the top panel of Figure~\ref{fig:hardness}, there is no clear offset in the PAH ratios in either panel at fixed \OIII/\hb\ between stellar clusters and diffuse regions.  The anti-correlation in F335M$_{\rm PAH}$/F1130W versus \OIII5007/\hb\ is consistent with the smallest PAHs being destroyed for regions with the hardest and most intense radiation fields (see also \citealt{lai2023}).  
The F770W$_{\rm dust}$/F1130W ratio in the bottom panel of Figure~\ref{fig:hardness} shows a much tighter decreasing trend than the trend observed in the top panel.  As described in \cite{baron2024}, this relationship is explicable if the ionized gas properties are determined by the ionizing portion of the radiation field and the PAH characteristics are driven by the non-ionizing component.

\begin{deluxetable}{llrcc}
\tabletypesize{\scriptsize}
\tablenum{2}
\tablecaption{PAH Feature Ratio Statistics \label{tab:medians}}
\tablewidth{0pc}
\tablehead{
\colhead{Galaxy} &
\colhead{Population} &
\colhead{Number} &
\colhead{med$_x$,med$_y$} &
\colhead{SIQR$_x$,SIQR$_y$} 
}
\startdata
\hline
NGC0628 & associations & 1379 & $-$1.02,$-$1.06 & 0.14,0.14\\
        & clusters     &  262 & $-$0.98,$-$1.01 & 0.14,0.14\\
        & diffuse      &   19 & $-$1.23,$-$1.30 & 0.24,0.19\\
NGC1087 & associations & 1023 & $-$0.81,$-$0.80 & 0.05,0.05\\
        & clusters     &  492 & $-$0.81,$-$0.80 & 0.06,0.06\\
        & diffuse      &   20 & $-$0.83,$-$0.82 & 0.06,0.04\\
NGC1300 & associations &  763 & $-$0.91,$-$0.90 & 0.10,0.11\\
        & clusters     &  182 & $-$0.91,$-$0.91 & 0.12,0.13\\
        & diffuse      &    8 & $-$1.20,$-$1.25 & 0.31,0.26\\
NGC1365 & associations &  622 & $-$1.04,$-$1.04 & 0.09,0.09\\
        & clusters     &  412 & $-$1.07,$-$1.10 & 0.09,0.09\\
        & diffuse      &   18 & $-$1.44,$-$1.48 & 0.22,0.26\\
NGC1385 & associations & 1012 & $-$0.91,$-$0.89 & 0.05,0.07\\
        & clusters     &  434 & $-$0.94,$-$0.93 & 0.08,0.08\\
        & diffuse      &   19 & $-$0.96,$-$0.98 & 0.09,0.09\\
NGC1433 & associations &  521 & $-$0.88,$-$0.91 & 0.12,0.13\\
        & clusters     &  147 & $-$0.85,$-$0.87 & 0.11,0.10\\
        & diffuse      &   12 & $-$1.08,$-$1.24 & 0.20,0.14\\
NGC1512 & associations &  697 & $-$0.80,$-$0.79 & 0.12,0.11\\
        & clusters     &  227 & $-$0.79,$-$0.82 & 0.15,0.12\\
        & diffuse      &   16 & $-$0.88,$-$0.96 & 0.11,0.11\\
NGC1566 & associations & 2503 & $-$0.99,$-$1.01 & 0.10,0.11\\
        & clusters     &  527 & $-$1.02,$-$1.06 & 0.12,0.13\\
        & diffuse      &   13 & $-$1.30,$-$1.35 & 0.22,0.18\\
NGC1672 & associations & 2842 & $-$0.96,$-$0.97 & 0.10,0.11\\
        & clusters     &  292 & $-$0.97,$-$0.96 & 0.08,0.08\\
        & diffuse      &   20 & $-$1.25,$-$1.34 & 0.24,0.23\\
NGC2835 & associations &  549 & $-$1.02,$-$1.02 & 0.18,0.18\\
        & clusters     &  210 & $-$1.00,$-$1.00 & 0.21,0.21\\
        & diffuse      &    3 & $-$1.65,$-$1.71 & 0.73,0.73\\
NGC3351 & associations &  543 & $-$1.09,$-$1.12 & 0.17,0.17\\
        & clusters     &  154 & $-$1.01,$-$1.05 & 0.15,0.14\\
        & diffuse      &    5 & $-$0.84,$-$1.08 & 0.89,0.75\\
NGC3627 & associations & 2785 & $-$0.85,$-$0.87 & 0.06,0.04\\
        & clusters     &  666 & $-$0.85,$-$0.88 & 0.07,0.04\\
        & diffuse      &   20 & $-$0.92,$-$0.93 & 0.09,0.06\\
NGC4254 & associations & 2854 & $-$1.04,$-$1.03 & 0.08,0.09\\
        & clusters     &  390 & $-$1.03,$-$1.05 & 0.07,0.09\\
        & diffuse      &   19 & $-$1.16,$-$1.18 & 0.12,0.12\\
NGC4303 & associations & 2134 & $-$1.09,$-$1.09 & 0.15,0.16\\
        & clusters     &  419 & $-$1.03,$-$1.04 & 0.15,0.16\\
        & diffuse      &    6 & $-$1.32,$-$1.38 & 0.19,0.19\\
NGC4321 & associations & 2127 & $-$1.01,$-$1.03 & 0.10,0.11\\
        & clusters     &  588 & $-$1.02,$-$1.04 & 0.10,0.10\\
        & diffuse      &   20 & $-$1.30,$-$1.40 & 0.16,0.10\\
NGC4535 & associations &  582 & $-$1.02,$-$1.01 & 0.12,0.14\\
        & clusters     &  219 & $-$1.07,$-$1.06 & 0.17,0.18\\
        & diffuse      &   18 & $-$1.36,$-$1.37 & 0.27,0.25\\
NGC5068 & associations &  882 & $-$0.81,$-$0.85 & 0.10,0.08\\
        & clusters     &  103 & $-$0.73,$-$0.81 & 0.10,0.06\\
        & diffuse      &   15 & $-$0.86,$-$0.98 & 0.07,0.03\\
NGC7496 & associations &  424 & $-$0.87,$-$0.85 & 0.07,0.08\\
        & clusters     &  217 & $-$0.89,$-$0.90 & 0.11,0.11\\
        & diffuse      &   20 & $-$1.00,$-$1.09 & 0.14,0.09\\
IC5332  & associations &  392 & $-$0.77,$-$0.74 & 0.21,0.23\\
        & clusters     &   90 & $-$0.75,$-$0.73 & 0.18,0.17\\
        & diffuse      &   16 & $-$0.85,$-$0.86 & 0.35,0.36\\
\hline
All     & associations & 24634 & $-$0.96,$-$0.96 & 0.11,0.12\\
        & clusters     &  6031 & $-$0.95,$-$0.96 & 0.12,0.12\\
        & diffuse      &   287 & $-$1.06,$-$1.11 & 0.20,0.19\\
\hline
\enddata
\tablenotetext{}{Statistics are based on sources with signal-to-noise ratios greater than 5 in all NIRCam and MIRI bands used in this analysis.  med$_x$ and med$_y$ refer to the medians of log$_{10}$(F335M$_{\rm PAH}$/F770W$_{\rm PAH}$) and log$_{10}$(F335M$_{\rm PAH}$/F1130W), respectively. The semi-interquartile ranges are half the 25\%--75\% range found after sorting the flux ratios.}
\end{deluxetable}

\section{Discussion} \label{sec:discussion}
Notwithstanding important efforts carried out with the {\it Infrared Space Observatory} and {\it Akari} (e.g., \citealt{moorwood1996,lai2020,kondo2024}), it is only with the recent advent of JWST and its exquisite sensitivity over its spectral coverage that we have been able to carry out detailed observations of PAH features that include the 3.35\m\ PAH feature (e.g., \citealt{peeters2024}).  Having access to observations of the smallest PAHs provides additional ways to understand the interplay between dust grains and the radiation fields that heat them.  The top two panels of Figure~\ref{fig:ratios.one.panel} show that older stellar clusters and associations correspond to, on average, lower F335M$_{\rm PAH}$/F770W$_{\rm dust}$ and F335M$_{\rm PAH}$/F1130W ratios.  In addition, the bottom panel of Figure~\ref{fig:ratios.one.panel} shows that diffuse regions clearly have, on average, lower F335M$_{\rm PAH}$/F770W$_{\rm dust}$ and F335M$_{\rm PAH}$/F1130W ratios than what are seen for stellar clusters and associations.  One reasonable interpretation is that dust grains and PAHs in diffuse regions are heated by photons originating from older stellar populations compared to the stellar populations driving the radiation fields in clusters and associations, consistent with expectations from theoretical models of PAH emission.  

However, comparing the median PAH feature ratios for each galaxy to the galaxy-averaged offsets from the star-forming main sequence and galaxy-averaged star formation rate surface densities (Figure~\ref{fig:offsets}) provide trends that are opposite to naive expectations.  One potential issue here are the limitations inherent in comparing localized phenomena to galaxy-wide properties.  But an additional piece of evidence points to a more nuanced intepretation: for 15 of the 19 of the galaxies in the sample the stellar clusters and diffuse regions show an anti-correlation when comparing F335M$_{\rm PAH}$/F1130W and the radiation field hardness tracer \OIII/\hb, and in fact the diffuse regions have on average slightly higher values of \OIII/\hb\ (Figure~\ref{fig:hardness}).  As described in \cite{belfiore2022}, the radiation escaping from density-bounded \HII\ regions into the surrounding diffuse gas can be harder than the radiation field within the \HII\ regions.  In this `filtered radiation' scenario the ionizing photons that are leaking from the \HII\ regions are harder because the absorption cross-section of hydrogen decreases with frequency $\nu$ as approximately $\nu^{-3}$.  The anti-correlations seen in Figure~\ref{fig:hardness} suggests that the PAH ratios are driven by radiation field hardness, for both stellar clusters and diffuse regions, and that the smallest PAHs that are captured by JWST F335M imaging are destroyed for the hardest radiation fields. 

Another consideration is the possible influence of the hot dust continuum for the two bands for which we have not explicitly accounted for such contamination---F770W and F1130W.  A significant hot dust continuum in the 7--12\m\ wavelength range would artificially depress the inferred 3.3\m/7.7\m\ and 3.3\m/11.3\m\ PAH feature ratios.  When comparing observed PAH feature ratios to model predictions, we have taken care to consistently account for stellar (and hot dust for F335M) continuum contributions to the observations and models.  However, the anti-correlation between the PAH feature ratios and \OIII/\hb\ is model-independent and serves as a key piece of evidence in our argument that PAH destruction impacts the observed F335M, F770W, and F1130W fluxes.  Figures~\ref{fig:ratios.one.panel.appendix}--\ref{fig:hardness.appendix} in the Appendix show the PAH feature ratio trends after removing estimates of the stellar and hot dust continua from the F770W and F1130W fluxes.

When PAH features appear to be relatively weak on $\gtrsim$100~pc scales in mid-infrared spectroscopy or photometry, the phenomenon is commonly ascribed to the destruction of PAHs due to hard radiation fields and other (sometimes entangled) environmental factors such as low metallicity, the local clearing of PAH dust by pre-supernova stellar feedback, or the effects of AGN \citep[e.g.,][]{houck2004,madden2006,sandstrom2012,chastenet2019,lai2022,zhang2024,whitmore2025}.  Though we have provided evidence for PAH destruction in our analysis, disentangling the various competing effects has proven difficult and ultimately requires i) a larger sample of galaxies to more fully explore parameters like metal abundance, ii) full mid-infrared spectroscopy, and iii) ideally the high spatial resolution afforded by nearby systems such as the Small and Large Magellanic Clouds and the Milky Way.  



\section{Conclusions} \label{sec:conclusions}
Mid-infrared PAH feature ratios are measured from JWST imaging data for 6\,031 stellar clusters and 24\,634 stellar associations for 19 nearby, solar metallicity, star-forming galaxies from the PHANGS project.  After smoothing to the coarsest resolution and careful removal of the underlying stellar continuum, the PAH feature ratios agree with the \cite{draine2021} model predictions, pointing to larger and more ionized PAH distributions.  The PAH feature ratios are also extracted for 287 diffuse regions spread throughout the galaxy disks to serve as a control for the cluster and association data.  The PAH feature ratios for the diffuse interstellar media in these galaxies are on average smaller in amplitude, seemingly consistent with the scenario of a softer and older stellar population driving the excitations of the PAH dust grains.  However, given the depressed PAH feature ratios for elevated \OIII/\hb\ ratios, for both clusters/associations and diffuse regions, our interpretation is that the PAH ratios are driven by the radiation field hardness, with the smallest PAHs traced by F335M imaging being destroyed in these harsh environments.  In the Appendix we show the effect of removing the residual hot dust continuum from the observed and synthetic model F770W and F11330W fluxes.  The trends are largely the same as presented in Section~\ref{sec:results}, except the anti-correlation in 7.7\m/11.3\m\ has a significantly shallower slope when the hot dust continuum is removed.  This shallower slope involving the two longer-wavelength PAH features---which generally trace larger PAHs---supports the notion that it is the smallest PAHs that are more likely to be destroyed.

To more securely quantify the contribution of PAH destruction to these mid-infrared PAH feature ratios, we will ultimately need the higher spatial resolution afforded by studies carried out in Magellanic Cloud and Milky Way environments \citep[e.g.,][]{cesarsky1996,chown2023}.  Exploring the PAH feature ratios for a wider range of metal abundances and star formation activity levels in nearby galaxies will also be critical.

This effort builds upon previous PAH feature ratio studies carried out on more limited galaxy samples \citep{chastenet2023a,dale2023,baron2024,ujjwal2024}.  Work in this area can be expanded upon in the future by leveraging the JWST Cycle~2 data for an additional 55 PHANGS galaxies.  Expanding to the larger PHANGS sample will also provide a larger range of metal abundances in which to explore these ratios.  In terms of the analysis carried out here, the main difference between PHANGS Cycle~1 and Cycle~2 datasets is the lack of F1130W imaging for Cycle~2 which will hamper interpretations that leverage characteristics of larger PAH grains.  However, as demonstrated in Figures~\ref{fig:ratios.one.panel}, \ref{fig:ratios.multi.panel}, \ref{fig:ratios.vs.ages}, and \ref{fig:offsets}, the PAH feature ratios F335M$_{\rm PAH}$/F770W$_{\rm dust}$ and F335M$_{\rm PAH}$/F1130W are well correlated and thus knowledge of this correlation from Cycle~1 data along with the measured F335M$_{\rm PAH}$/F770W$_{\rm dust}$ ratios from Cycle~2 data should enable useful diagnostics of PAH size and ionization level.


\section{acknowledgments}
We thank the referee for their helpful recommendations.
The PHANGS data presented in this article were obtained from the Mikulski Archive for Space Telescopes (MAST) at the Space Telescope Science Institute (STScI) and from Canadian Advanced Network for Astronomical Research (CANFAR). The HST and JWSTS specific PHANGS observations analyzed can be accessed via \dataset[doi: 10.17909/t9-r08f-dq31]{https://doi.org/10.17909/t9-r08f-dq31} and \dataset[doi: 10.17909/jray-9798]{https://doi.org/10.17909/jray-9798}, respectively. The CANFAR data are available at \dataset[doi: 10.11570/22.0082]{http://doi.org/10.11570/22.0082}.
This work is based on observations made with the NASA/ESA/CSA JWST and Hubble Space Telescopes. The data were obtained from MAST at STScI, which is operated by the Association of Universities for Research in Astronomy, Inc., under NASA contract NAS 5-03127 for JWST and NASA contract NAS 5-26555 for HST. The JWST observations are associated with Program~2107, and those from HST with Program~15454.
D.D. acknowledges support from grant JWST-GO-02107.005-A.
A.W. acknowledges UNAM's DGAPA for the support in carrying out her sabbatical stay at UCSD through program PASPA.
RSK and S.C.O.G. acknowledge funding from the European Research Council via the ERC Synergy Grant ``ECOGAL'' (project ID 855130), from the German Excellence Strategy via the Heidelberg Cluster of Excellence ``STRUCTURES'' (EXC 2181 - 390900948), and from the German Ministry for Economic Affairs and Climate Action in project ``MAINN'' (funding ID 50OO2206). 
R.S.K. and S.C.O.G. also thank for computing resources provided by {\em The L\"{a}nd} and DFG through grant INST 35/1134-1 FUGG and 35/1597-1 FUGG, and for data storage at SDS@hd through grant INST 35/1314-1 FUGG and INST 35/1503-1 FUGG. 
R.S.K. also thanks the Harvard Radcliffe Institute for Advanced Studies and the Harvard-Smithsonian Institute for Astrophysics for their warm welcome and hospitality during a sabbatical year. 
E.R. acknowledges support from the Canadian Space Agency, funding reference 23JWGO2A07.
M.B. gratefully acknowledges support from the ANID BASAL project FB210003 and from the FONDECYT regular grant 1211000. This work was supported by the French government through the France 2030 investment plan managed by the National Research Agency (ANR), as part of the Initiative of Excellence of Université Côte d’Azur under reference number ANR-15-IDEX-01.


\newpage
\appendix
Though we have taken care to remove the stellar and hot dust continua for the (observed and modeled) F335M fluxes, and the stellar continuum for the F770W fluxes, there may also be significant hot dust continuum contributions to the F770W and F1130W bandpasses.  For example, \cite{whitcomb2023} show that the dust continuum contribution over $>$100 pc scales in local star-forming galaxies is $\sim$15\% for F770W and $\sim$25\% for F1130W.  The PHANGS team has developed a new (stellar+dust) continuum subtraction recipe (Hands et al., in preparation) that is based on Spitzer Space Telescope mid-infrared spectra of nearby star-forming galaxies coupled with synthetic F770W/F1000W/F1130W photometry, similar to what was done for \cite{whitcomb2023}.  We find results comparable to those described in \cite{whitcomb2023}: the typical continuum-free PAH fractions are 83\% and 75\% for F770W and F1130W, respectively.  

Similar to how we refer to the stellar and hot dust continuum-subtracted 3.3\m\ PAH fluxes as F335M$_{\rm PAH}$, in Figures~\ref{fig:ratios.multi.panel.appendix}--\ref{fig:hardness.appendix} we refer to the stellar and hot dust continuum-subtracted 7.7\m\ and 11.3\m\ PAH fluxes as F770W$_{\rm PAH}$ and F1130W$_{\rm PAH}$.

\begin{figure}
 \includegraphics[height=16.5cm]{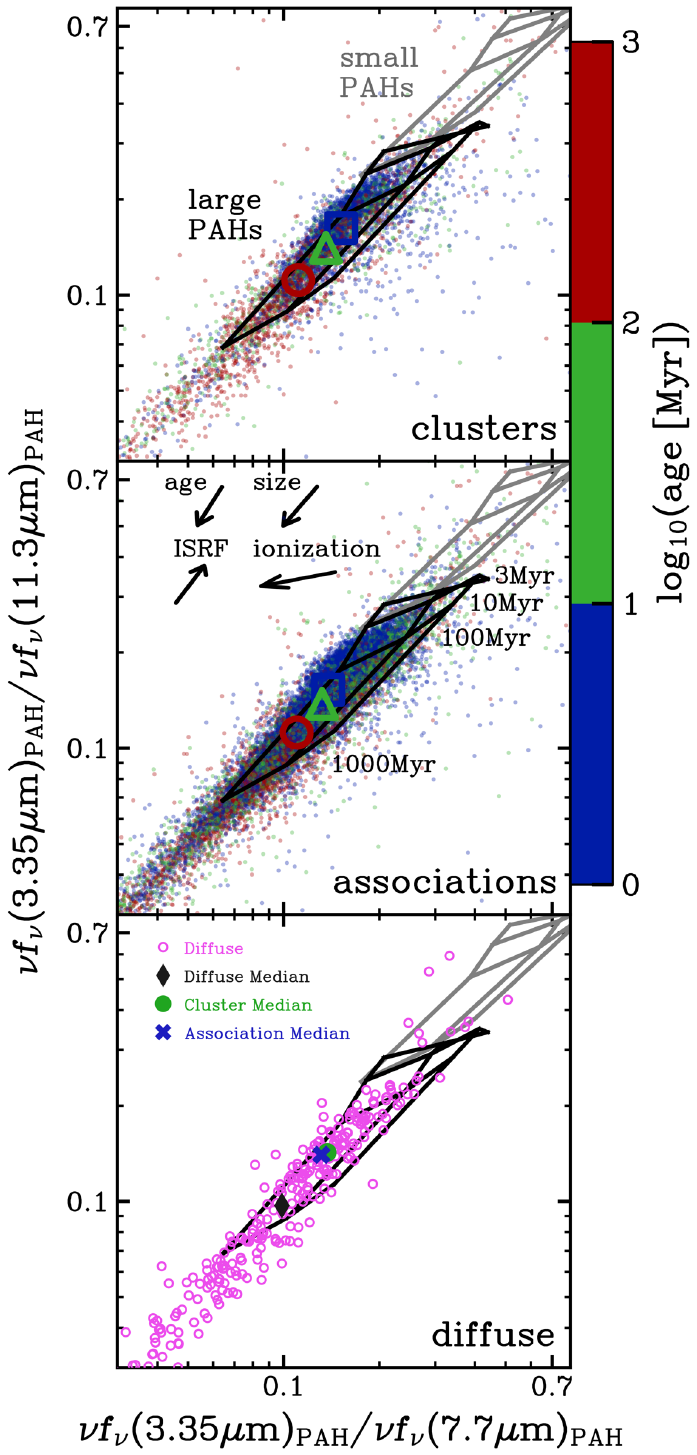}
 \caption{Similar to Figure~\ref{fig:ratios.one.panel} but after removing the hot dust continuum levels for F770W and F1130W.}
 \label{fig:ratios.one.panel.appendix}
\end{figure}

\begin{figure}
 \includegraphics[height=16.5cm]{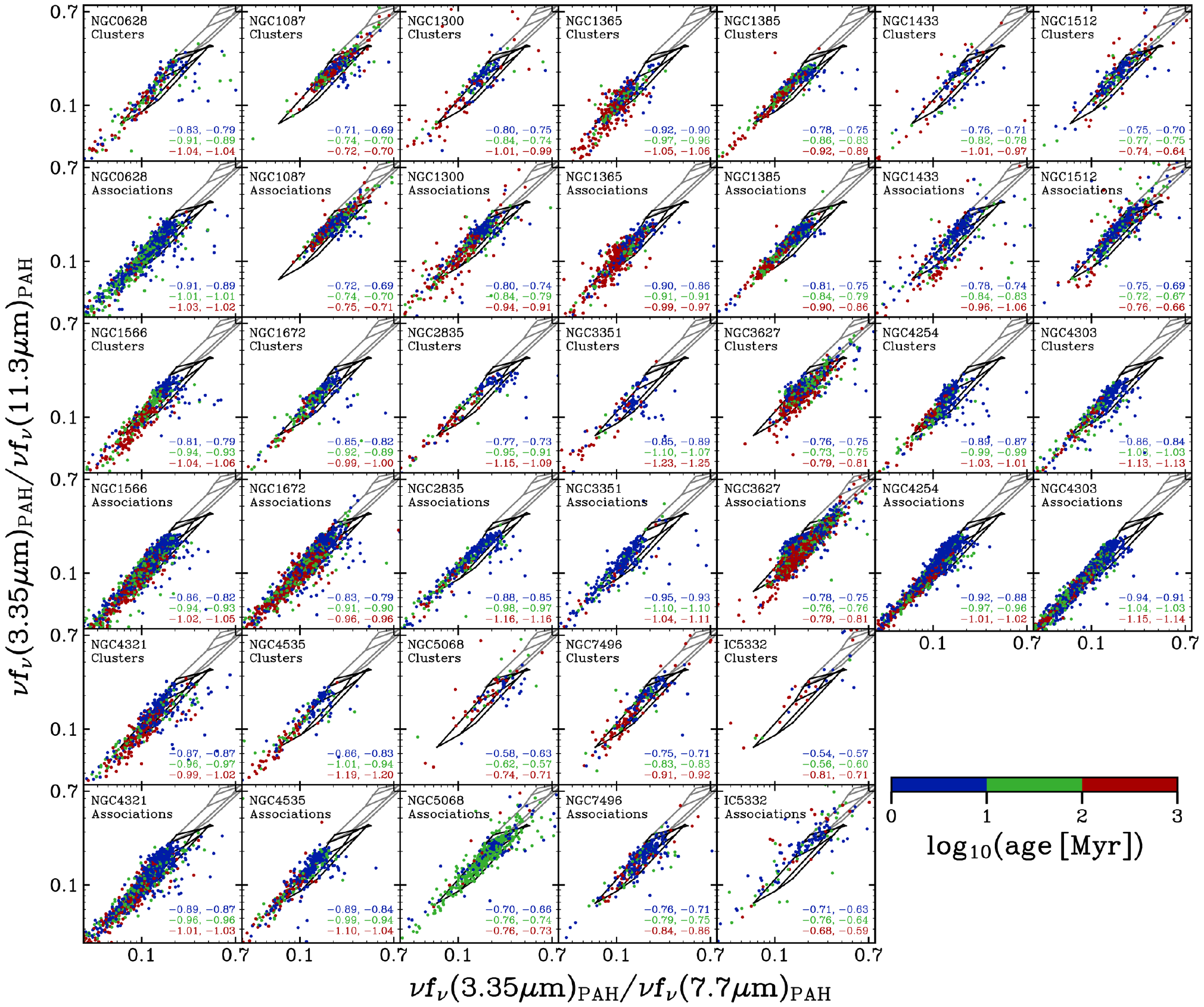}
 \caption{Similar to Figure~\ref{fig:ratios.multi.panel} but after removing the hot dust continuum levels for F770W and F1130W.}
 \label{fig:ratios.multi.panel.appendix}
\end{figure}

\begin{figure}
 \includegraphics[height=8cm,clip]{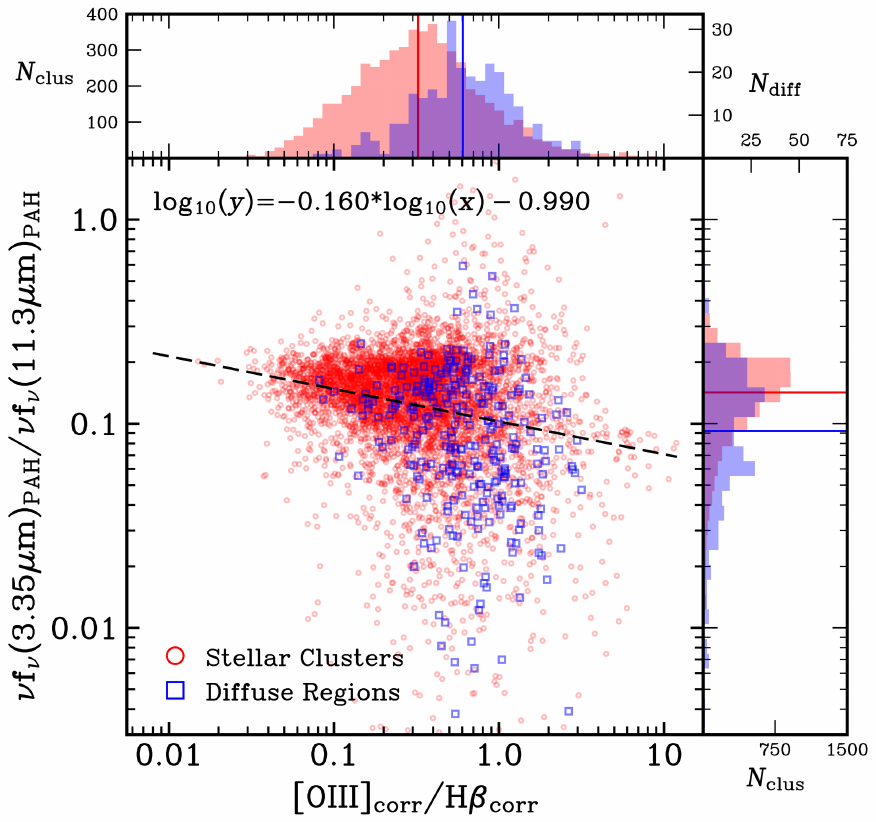}
 \includegraphics[height=8cm,clip]{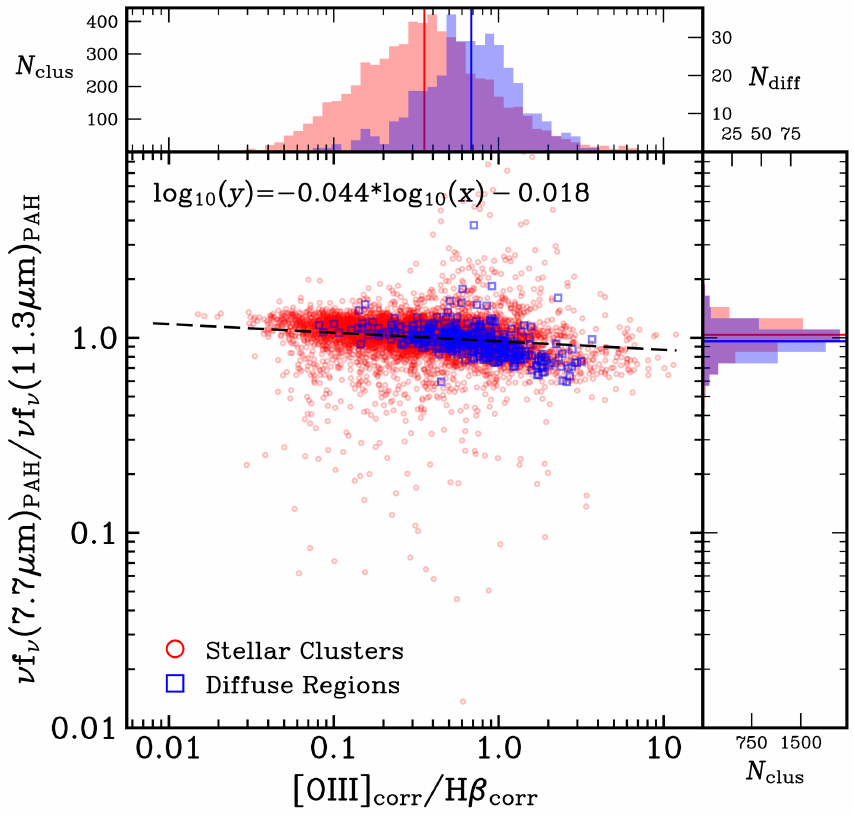}
 \caption{Similar to Figure~\ref{fig:hardness} but after removing the hot dust continuum levels for F770W and F1130W.}
 \label{fig:hardness.appendix}
\end{figure}

\bibliography{main.bib}




\end{document}